\documentclass[12pt]{report}
\usepackage{epsf,amsmath}
\usepackage{indentfirst}
\usepackage{amssymb}
\usepackage{array,tabularx}
\begin{document}

\title{\large \bf The Mass of the Compact Object in the X-Ray Binary Her X-1/HZ Her
}

\author{\large\bf M. K. Abubekerov$^1$, E. A. Antokhina $^1$, A. M. Cherepashchuk$^1$,\\
\large\bf V. V. Shimanskii$^2$ \\
\normalsize\it $^1$ Sternberg Astronomical Institute, Russia \\
\normalsize\it $^2$ Kazan State University, Russia \\
}

\date{\begin{minipage}{15.5cm} \small
We have obtained the first estimates of the masses of the
components of the Her X-1/HZ Her X-ray binary system taking into
account non-LTE effects in the formation of the $\text H_{\gamma}$
absorption line: $m_x=1.8M_{\odot}$ and $m_v=2.5M_{\odot}$. These
mass estimates were made in a Roche model based on the observed
radial-velocity curve of the optical star, HZ Her. The masses for
the X-ray pulsar and optical star obtained for an LTE model lie
are $m_x=0.85\pm0.15M_{\odot}$ and $m_v=1.87\pm0.13M_{\odot}$.
These mass estimates for the components of Her X-1/HZ Her derived
from the radial-velocity curve should be considered tentative.
Further mass estimates from high-precision observations of the
orbital variability of the absorption profiles in a non-LTE model
for the atmosphere of the optical component should be made.
\end{minipage}
} \maketitle \rm

\section*{\normalsize INTRODUCTION}

The X-ray source Her X-1 was detected by the Uhuru satellite in
1972 \cite{Tananbaum1972} and identified with the variable star HZ
Her \cite{Kurochkin1972, Cherepashchuk1972,
Forman1972,Bahcall1972}. The Her X-1/HZ Her binary system consists
of the A7 optical component \cite{Cheng1995}, which fills its
Roche lobe, and an X-ray pulsar with a period of 1.24 s. The
orbital period of the binary is Porb = 1.7d \cite{Prince1994}. The
system also displays 35-day X-ray variability due to the
precession of the warped accretion disk
\cite{Tananbaum1972,Schandl1996}. The optical variability of the
binary is mainly due to X-ray heating of the optical star
\cite{Cherepashchuk1972,Bahcall1972}.

The Her X-1/HZ Her binary has been thoroughly studied: the
high-precision half-amplitude of the radial-velocity curve for the
X-ray pulsar $K_x=169.049$ km/s \cite{Deeter1981} has been derived
from X-ray timing, and the orbital inclination
$i=81^{\circ}-88^{\circ}$ \cite{Cheng1995,Reynolds1997} from the
duration of the X-ray eclipses. However, no unambiguous estimates
for the mass of the X-ray pulsar have been obtained, due to the
very intense X-ray heating, $k_x\simeq150$.

The value of a neutron star in the Her X-1/HZ binary is not
determined unambiguously. According to Middleditch and Nelson
\cite{Middleditch1976}, the mass of the X-ray pulsar is
$m_x\simeq1.30\pm0.14M_{\odot}$. Koo and Kron \cite{Koo1977}
suggest that the mass of the compact object in Her X- 1 is close
to $m_x\simeq1.5M_{\odot}$. The mass estimated by Hutchings et al.
\cite{Hutchings1985} is $m_x=0.93\pm0.07M_{\odot}$, while Reynolds
et al. \cite{Reynolds1997} conclude this mass lies in the interval
$m_x=1.5\pm0.3M_{\odot}$. These estimates were made using
point-mass models, except for the estimate of Koo and Kron
\cite{Koo1977}, which was made for a Roche model assuming a
spherical shape for the optical component and roughly taking into
account heating of this component.

The difficulty in accurately estimating the pulsar mass is
essentially due to the intense X-ray heating of the optical
component. The X-ray luminosity of the compact object is
$L_x=3.39\cdot10^{37}$ erg/s \cite{Cheng1995}, causing the
spectral type of the optical star to vary from A7 ($\phi=0.0$) to
B3-B6 ($\phi=0.5$) over half the orbital cycle
\cite{Cheng1995,Anderson1994}. The temperature distribution over
the surface of the optical star is also complicated by shielding
of the X-ray radiation by the warped, precessing accretion disk.
Together, these factors hinder efforts to measure a
high-precision, purely orbital radial velocity curve, and thereby
derive an unambiguous estimate for the mass of the X-ray pulsar.

Note that, in a point-mass model, only the halfamplitude of the
radial-velocity curve, $K_v$, is taken into account, not its
shape. However, it is known that the observed radial-velocity
curves of stars in binary systems may differ from the
radial-velocity curves for the centers of mass, due to
tidal–rotational distortion of the stellar surfaces and to heating
by the radiation from their companions (see the studies of Wilson
and Sofia \cite{Wilson1976}, Milgrom \cite{Milgrom1977}, and
references therein). It is particularly important to take these
effects into account when determining themasses of relativistic
objects in X-ray binaries. For example,Milgrom
\cite{Milgrom1976,Milgrom1977} showed that, even in the case of
relatively low X-ray heating of the optical star (as in Cyg X-1
and SMC X-1), the radial velocities derived from absorption lines
can vary substantially. Using profile calculations for various
absorption lines in the spectra of HZ Her and other stars
\cite{Milgrom1976,Milgrom1977}, he noted the important role of the
"soft" X-ray radiation ($E<1$ keV) in the spectrum of the
relativistic object, which is responsible for the emission
components of the lines. Milgrom and Salpeter \cite{Milgrom1975}
developed a procedure for calculating the radiation emerging from
the atmosphere of a star irradiated by X-ray radiation, intended
for the calculation of line profiles. Unfortunately, in the 1970s,
such calculations required enormous amounts of computer time, and
could not be applied to the massive calculations required to solve
this problem.

Antokhina et al. \cite{Antokhina2003,Antokhina2005} suggested a
procedure for calculating line profiles and radial-velocity curves
of optical stars in X-ray binaries taking into account the
incident X-ray flux from the relativistic object. Using this
technique, we carried out a series of studies analyzing
spectroscopic observations of close X-ray binaries
\cite{Abubekerov2004a,Abubekerov2004b,Abubekerov2005,Abubekerov2006},
and demonstrated the need to take into account the shape of the
radial-velocity curve when determining the component masses,
particularly in the presence of intense X-ray heating.

Here, we analyze the observational data of \cite{Reynolds1997} in
a Roche model exactly taking into account the effect of X-ray
heating of the optical star
\cite{Antokhina1994,Antokhina1996,Antokhina2005}, with the aim of
reliably estimating the mass of the compact object in Her X-1/HZ
Her. Running ahead for a moment, we note that no emission
components are visible in the absorption lines in the spectrum of
the optical star HZ Her, despite the extremely intense X-ray
heating. Since this may be related to deviations from LTE, we have
analyzed the radial-velocity curve using both LTE and non-LTE
models.

\section*{\normalsize OBSERVATIONS}

Let us briefly describe the spectral data adopted from
\cite{Reynolds1997}. The observations were carried out on June
10-16, 1995 with the 2.5-m Isaac Newton Telescope of the Royal
Greenwich Observatory at La Palma (Canary Islands). In total, 59
spectra at wavelengths of 4080-4940\,\AA\AA\, and with exposures
from 1200-1800 secunds were obtained with the Intermediate
Dispersion Spectrograph equipped with a 235-mm camera, a
diffraction grating with a dispersion of 1200 lines/mm, and a
1024x1024 pix Tektronix CCD array. The inverse linear dispersion
of the spectrograms was 0.84\,\AA/pixel. The signal-to-noise
ratios in the spectra obtained were $S/N=50-100$ The wavelength
scale was calibrated with a Cu-Ar standard; the calibration error
was $<0.05$ \AA.

The radial velocity of HZ Her was determined via a
cross-correlation analysis using the radial velocities of IAU
standard F stars observed on the same nights as HZ Her. The epoch
of the maximum positive radial velocity of the optical component
(JD=2448800.537) was used as the zero phase in
\cite{Reynolds1997}. We adopted the orbital period
$P_{orb}=1^{d}.700167412$ \cite{Prince1994}.

\begin{figure*}
\vspace{0cm} \epsfxsize=0.99\textwidth
\epsfbox{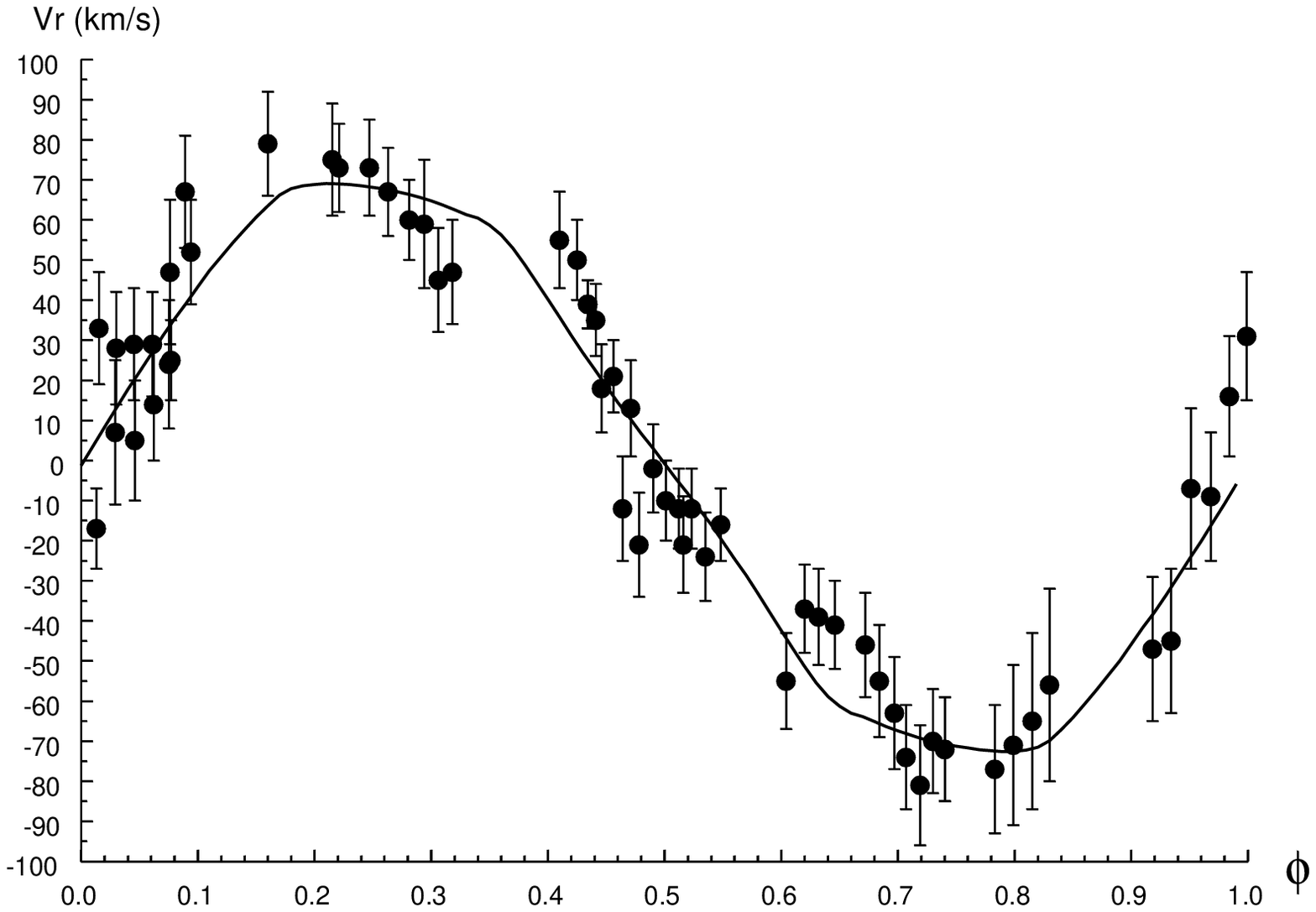}
\caption{Observed and theoretical radial-velocity curves of the
optical component of the X-ray binary Her X-1/HZ Her. The points
indicate the radial velocities of HZ Her taken from
\cite{Reynolds1997}. The solid curve is the theoretical
radial-velocity curve for a Roche model calculated using Procedure
II, with the mass of the compact object $m_x=0.81M_{\odot}$, the
mass of the optical star $m_v=1.80M_{\odot}$, and the orbital
inclination $i=88^{\circ}$. (for the remaining parameters of the
binary, see Table \ref{param Roche}).} \label{Vobs}
\end{figure*}

Here, we took the zero orbital phase to correspond to the middle
point of the eclipse of the X-ray source by the optical component.
The data analysis of \cite{Reynolds1997} shows that, in this case,
the maximum positive radial velocity of the optical star falls at
phase $\phi=0.22$, and the systematic radial velocity of the
center of mass is $\gamma=-69$ km/s. Figure \ref{Vobs} presents
the observed radial-velocity curve.

The spectrograms of \cite{Reynolds1997} were obtained for the
orbital-phase interval of the 35-day precessional cycle of the
accretion disk $\phi_{35}=0.14-0.33$ (the zero phase was ascribed
to the moment when the X-ray source "switches on" for an observer
on Earth). Thus, at the epoch of the spectroscopical observations
\cite{Reynolds1997}, the accretion disk did not shield the optical
star from the X-ray flux, and our fitting of the observed
radial-velocity curve did not take into account this shielding.

\section*{\normalsize MODEL OF THE BINARY
AND PROCEDURE FOR CALCULATING THE RADIAL-VELOCITY CURVE}

We synthesized theoretical absorption-line profiles and
radial-velocity curves for the optical star using the two
procedures described in detail in
\cite{Antokhina1994,Antokhina1996,Antokhina2005}, which we call
Procedure I \cite{Antokhina1994,Antokhina1996} and Procedure II
\cite{Antokhina2003,Antokhina2005}. These procedures are similar,
and differ only in the means used to calculate the radiation flux
from local areas of the star. Procedure I (the more rapid
algorithm) uses the Balmer line profiles tabulated by Kurucz
\cite{Kurucz1979}; in Procedure II (the more accurate algorithm),
the emergent flux is calculated using a model atmosphere in the
presence of incident external flux. We will briefly describe the
calculation technique.

In a close binary, the optical component has a tidally distorted
shape and an inhomogeneous temperature distribution over its
surface, due to the effects of gravitational darkening and surface
heating by the X-ray radiation from the relativistic object.
Therefore, to model the radial-velocity curve, the X-ray binary
was represented by the optical star in the Roche model and an
X-ray point source with a finite mass. The tidally deformed
stellar surface was divided into $\sim2600$ elements, for each of
which the emergent radiation was calculated. The calculations of
the flux from a surface element included gravitational darkening,
the heating of the stellar surface by radiation from its companion
(the reflection effect), and limb darkening. In Procedure I
\cite{Antokhina1994,Antokhina1996}, the heating of the star's
atmosphere by X-ray radiation from the companion was calculated by
adding the emergent and incident flux without considering
radiative transfer in the stellar atmosphere.

For each visible surface element with temperature $T_{loc}$ and
local gravitational acceleration $g_{loc}$, the absorption-line
profiles and their equivalent widths were interpolated from the
Kurucz tables for Balmer lines \cite{Kurucz1979}. Summing the
local profiles over the visible surface of the star, taking into
account the Doppler effect, and normalizing the profiles to the
continuum level for each surface element, we calculated the
integrated profile from the star at a given orbital phase (see
\cite{Antokhina1994,Antokhina1996} for details). The calculated
integrated absorption-line profile was used to determine the
radial velocity of the star. The radial velocity at a given
orbital phase was calculated for the average wavelength at the
levels of the residual intensities at 1/3, 1/2, and 2/3 of the
integrated absorption profile.

In the more modern Procedure II, for each surface element, the
parameter $k_{x}^{loc}$, equal to the ratio of the incident X-ray
flux and emergent radiation flux, is calculated separately from
the local temperature $T_{loc}$ and local gravitational
acceleration $g_{loc}$, without allowing for the external
irradiation of the atmosphere. The model atmosphere is calculated
for these parameter values at a given point of the surface,
assuming LTE, by solving the equations of radiative transfer in
the line in the presence of incident external X-ray radiation (see
\cite{Antokhina2003,Antokhina2005} for details). Thus, the
intensity of the emergent radiation in the line and continuum is
calculated for each local surface element. Further, following a
procedure identical to Procedure I, the integrated absorption-line
profile at the given orbital phase is calculated, which is then
used to determine the radial velocity of the optical star.

In summary, the main difference between Procedure II
\cite{Antokhina2003,Antokhina2005} and Procedure I
\cite{Antokhina1994,Antokhina1996} is that, in the former the
local profile of a surface element is found by constructing an LTE
model of the atmosphere and calculating the intensities of
emergent radiation in the line and in continuum (taking into
account reprocessing of the external X-ray radiation). In
contrast, Procedure I uses calculated and tabulated Kurucz Balmer
absorption-line profiles for various effective temperatures
$T_{eff}$ and gravitational accelerations $g$. Note that, in
addition to the simplified treatment of the reflection effect,
this method for calculating the hydrogen absorption profiles is
not entirely correct, since the theoretical line profiles in
Kurucz's tables \cite{Kurucz1979} are given in relative fluxes
rather than intensities. However, since we use the theoretical
line profiles to determine radial velocities and not for
comparison with the observed profiles, we consider this
approximation to be reasonable \cite{Abubekerov2005}.


Substantial deviations from LTE can occur in the atmospheres of
hot stars, particularly in the presence of external irradiation.
Therefore, we modeled the Balmer line profiles with Procedure II
both assuming LTE and taking into account non-LTE effects in the
hydrogen atom. The equilibrium non-LTE populations of the HI
levels were calculated using a technique similar to that developed
by Ivanova et al. \cite{ivan1}. In the calculations, a 23-level
model of the $HI$ atom was applied, taking into account all
allowed collisional and radiative mechanisms for the atom's
redistribution over the states, including those generated by the
external radiation. In total, the $HI$ model includes 153 allowed
bound-bound and 22 bound-free radiative transitions, 55 of which
were linearized. The non-LTE populations of the levels were
obtained for stellar-atmosphere models for the local surface
elements using the $NONLTE3$ \cite{sahi2} package, which
implements the total linearization technique of Auer and Heasley
\cite{auer1}, modified by Ivanova et al. \cite{ivan1} to take into
account the external radiation. The calculations included all
continuum sources of opacity at optical and X-ray wavelengths and
about 570 000 lines from the lists of Kurucz \cite{kur1}. The
subsequent modeling of HI line profiles was carried out following
the above standard technique, but using the non-LTE populations.

\begin{figure*}
\vspace{0cm} \epsfxsize=0.99\textwidth
\epsfbox{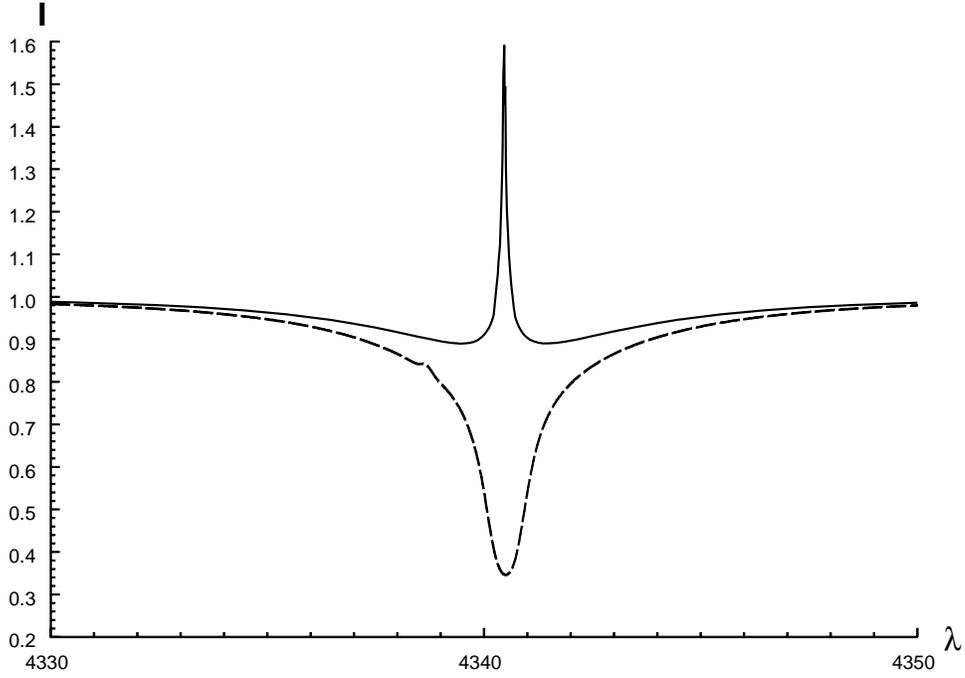} \caption{Model
local profiles for the $\text H_{\gamma}$ absorption line. The
solid curve corresponds to the calculations assuming LTE
(Procedure II) for $m_x=0.9M_{\odot}$, $m_v=2.0M_{\odot}$,
$i=80^{\circ}$, $k_x=150$, and the dashed curve to the non-LTE
calculations for the same parameters of the binary. The
temperature and gravitational acceleration for the local surface
element are $T_{loc}= 7825$K and $\log g_{loc} = 3.336$.}
\label{Hg_loc}
\end{figure*}

When the local profiles in Fig.\ref{Hg_loc} obtained for the LTE
and non-LTE models are compared, it is obvious that the $\text
H_{\gamma}$ line does not display an emission component in the
non-LTE case. This is due to the strong (by a factor of four to
six) overpopulation of the second level ($n=2$) of the hydrogen
atom (relative to the above levels) in the chromosphere. The
source function $S_\nu \simeq \frac{b_j}{b_i} B_\nu (T_e)$ is
proportional to the ratio of the populations of the lower and
upper levels of the transition. As a result, in spite of the
substantial increase of $B_\nu (T_e)$ in the hot chromosphere,
$S_\nu$ exceeds the source function in the region of formation of
the continuum; i.e., the line profile appears in absorption.

As follows from an analysis of physical processes in the $HI$
atom, an overpopulation of its low-excitated states should be
characteristic of hot objects with Xray irradiation. According to
\cite{sahi1}, hydrogen is responsible for less than $20\%$ of the
total absorption at X-ray wavelengths for $T_e=7000K$, and for a
factor of $10^3$ less for $T_e=20000K$.

At the same time, since the temperature in the chromosphere
exceeds that in the region of continuum formation, the source
function in the chromosphere dominates over the mean intensity of
the radiation at all optical frequencies, $J_\nu \ll S_\nu$, where
$J_\nu$ is the mean intensity of the radiation and $S_\nu$ the
total source function, both at a given frequency at a given depth.
Thus, radiative recombination to all $HI$ levels and subsequent
spontaneous transitions, which are proportional to $S_\nu$,
dominate over ionization processes due to both the external
radiation and intrinsic radiation of the atmosphere. As a result,
all $HI$ levels turn out to be overpopulated relative to the LTE
approximation, with the overpopulation increasing for lower
states.

These non-LTE effects should result in a strengthening of the
absorption and weakening of the emission components of hydrogen
lines in the spectrum of an atmosphere with substantial X-ray
irradiation. We wish to emphasize that, in the spectrum of the
optical star HZ Her, in spite of the huge amount of X-ray heating
$(k_x\simeq 150)$, no emission components are observed in the
profiles of absorption lines, possibly due to deviations from LTE.

\section*{\normalsize ANALYSIS OF THE OBSERVED
RADIAL-VELOCITY CURVE}

We fitted the observed radial-velocity curve of the optical star
within the Roche model. Table \ref{param Roche} presents the model
parameters for the X-ray binary Her X-1/HZ Her.

\begin{table}[h!]
\caption{Parameters used in the modeling of the radial-velocity
curves of the optical component in the Roche model.}\label{param
Roche} \vspace{3.0mm} \centering
\def\arraystretch{1.0}
\newcolumntype{C}{>{\centering\arraybackslash}m}
\begin{tabular}{|l|l|p{110mm}|}
\hline
$P$(ñóò.)         & 1.700167412      & Orbital period \\
$e$               & 0.0              & Eccentricity \\
$i(^\circ)$       & $80$; $88$       & Orbital inclination \\
$\mu$             & 1.0              & Degree of Roche lobe filling by the optical component \\
$f$               & 1.0              & Asynchronicity factor for the rotation of the optical component \\
$T_{\text{eff}}$(K)& 8100            & Effective temperature of the non-perturbed optical component \\
$\beta$           & 0.08             & Gravitational-darkening coefficient \\
$k_x$             & 150              & Ratio of the X-ray
                                       luminosity of the relativistic component
                                       and the bolometric luminosity of the optical component, $L_x/L_{v}$ \\
$A  $             & 1.0              & Coefficient for reprocessing of the X-ray radiation (Procedure I) \\
$u  $             & $0.3$            & Limb-darkening coefficient \\
\hline
\end{tabular}
\end{table}

As is noted above, Her X-1/HZ Her displays strong X-ray heating of
the atmosphere of the optical star. The X-ray luminosity of the
compact object is $L_x\simeq3.39\cdot 10^{37}$ erg/s
\cite{Cheng1995}, whereas the luminosity of the optical star is
$L_{v}\simeq2.28\cdot10^{35}$ erg/s (the optical luminosity was
obtained in the Roche model assuming the mean effective
temperature $T_{eff}=8100$K \cite{Cheng1995}).

\begin{figure*}
\vspace{0cm} \epsfxsize=0.99\textwidth
\epsfbox{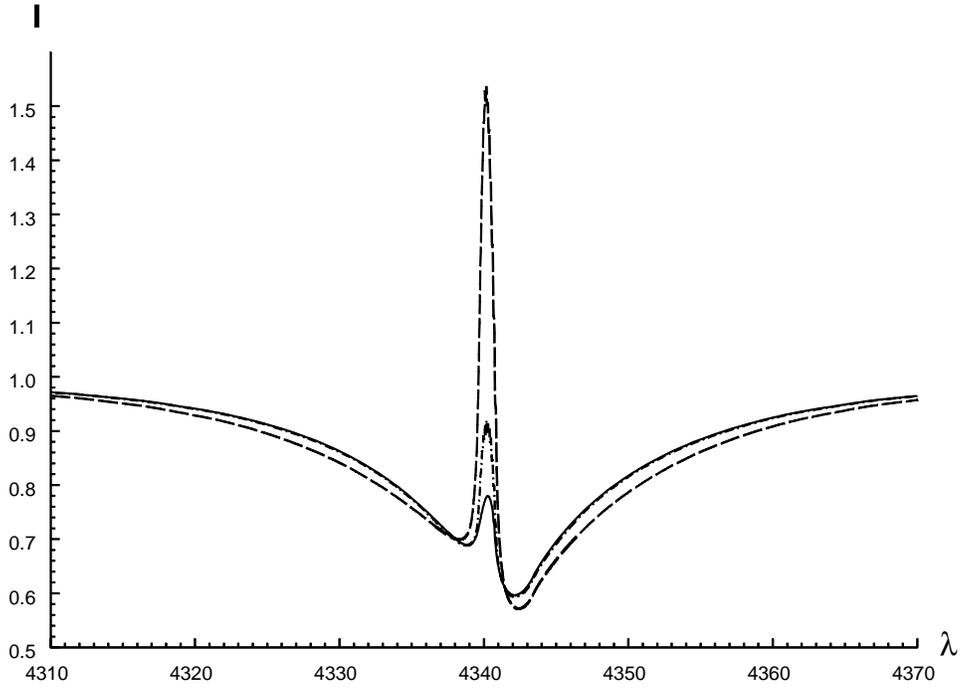}
\caption{Effect of taking into account the soft component of the
incident X-ray flux on the shape of the $\text H_{\gamma}$ line.
Integrated profiles calculated with Procedure II for
$m_x=1M_{\odot}$, $m_v=2M_{\odot}$, $i=80^{\circ}$, $k_x=150$ are
shown. The X-ray irradiation was modeled at 0.1-1 keV(dashed
curve), 0.1-17 keV (dash.dotted curve), and 1-17 keV (solid
curve).} \label{Hg_SpXray}
\end{figure*}

In our modeling of the heating of the atmosphere of the optical
component, we used the X-ray spectrum of Her X-1 obtained with
BeppoSAX in 1998 \cite{Oosterbroek1997}. In spite of the strong
interstellar absorption, soft X-ray radiation from Her X-1 at
0.1-1 keV is reliably detected, which is essential to study the
conditions for the formation of the optical spectrum of HZ Her
\cite{Milgrom1976}. For example, Fig.\ref{Hg_SpXray} presents two
model LTE spectra of the $\text H_{\gamma}$ hydrogen absorption
line of the optical component calculated with and without taking
into account the soft component (0.1-1 keV) of the irradiating
X-ray spectrum. When this soft component is taken into account,
the emission component in the $\text H_{\gamma}$ absorption line
is appreciably stronger.

The X-ray spectrum at 0.1-10 keV, corrected for the interstellar
absorption, was reconstructed from the data of
\cite{Oosterbroek1997} using the XSPEC code. The two-component
model of the spectrum was taken from \cite{Oosterbroek1997}:
blackbody radiation with the temperature 0.093 keV (which
dominates up to $\sim 1$ keV), and power-law-like radiation with
spectral index 0.74 (which dominates at energies exceeding 1 keV).
We did not include the components from \cite{Oosterbroek1997} that
describe the X-ray iron emission lines due to their small
contribution to the total X-ray flux.Whenmodeling the optical
spectrum of HZ Her, we assumed the X-ray radiation of the
relativistic object to be isotropic.

Since the exact mass of the optical star in the Her X-1/HZ Her
system is not known, we adopted the masses of both components
$m_x$, $m_v$ as parameters to be fit. We obtained the fit via an
exhaustive search over the parameters, making it possible to study
the surface of residuals in terms of various parameters in detail.

The half-amplitude of the radial-velocity curve of the X-ray
pulsar, $K_x=169.049$ km/s \cite{Deeter1981}, is known from X-ray
timing, and consequently the mass function of the pulsar is
$f_x(m)=0.85M_{\odot}$. Therefore, in the course of the exhaustive
search over the component masses, we particularly made sure that
the mass function of the X-ray pulsar retained the value
$f_x(m)=0.85M_{\odot}$. This was done as follows.

The mass function

\begin{equation}
f_x(m)=\frac{m_v^3\sin^3i}{(m_x+m_v)^2}. \label{fmx}
\end{equation}

\noindent for a given orbital inclination $i$ specifies an
unambiguous relation between the masses of the optical star $m_v$
and the X-ray pulsar $m_x$.

A value of $m_x$ was calculated from (\ref{fmx}) for each value of
$m_v$ and a fixed value of $i$. Further, the theoretical
radial-velocity curve for the optical component was determined
based on the resulting component masses $m_x$ and $m_v$, with the
numerical parameters of the Roche model taken from Table
\ref{param Roche}. The adequacy of the model to the observational
data was tested using the statistical $\chi^2$ criterion,
selecting a significance level of 5\% (see
\cite{Cherepashchuk1993} for details).

\section*{\normalsize\it Results Obtained with Procedure I}

\begin{figure*}
\vspace{0cm} \epsfxsize=0.99\textwidth
\epsfbox{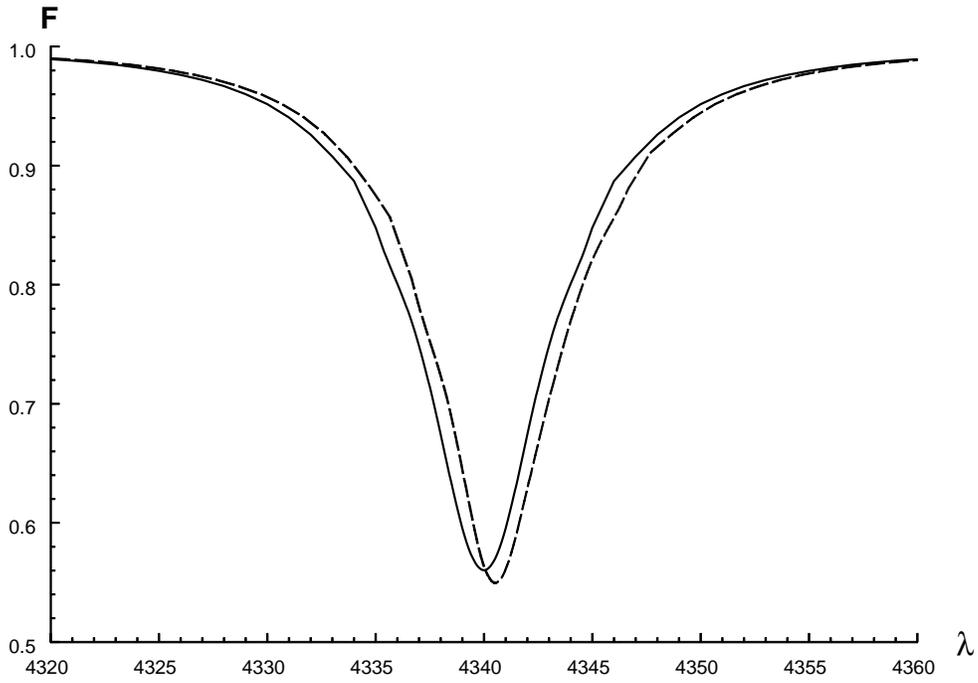}
\caption{Integrated profiles of the $\text H_{\gamma}$ absorption
line calculated in the Roche model with Procedure I, for the width
of the response function FWHM=1.7\,\AA and with
$m_x=1.78M_{\odot}$, $m_v=2.5M_{\odot}$, and $i=88^{\circ}$. The
solid line corresponds to the orbital phase $\phi=0.0$, and the
dashed line to $\phi=0.30$.} \label{Hg_Roche_old}
\end{figure*}

Since the radial velocities of \cite{Reynolds1997} were derived
from spectra obtained with the response function FWHM=1.7\,\AA, we
calculated the theoretical radial-velocity curves from the profile
of the $\text H_{\gamma}$ line convolved with an instrumental
profile with FWHM=1.7\,\AA. Figure \ref{Hg_Roche_old} presents the
theoretical integrated profiles of the $\text H_{\gamma}$
absorption line.

\small
\begin{table}[h!]
\caption{Mass of the relativistic and optical components of Her
X-1 obtained in the Roche model with Procedure
I.}\label{old_model} \vspace{3.0mm}\centering
\def\arraystretch{1.4}
\newcolumntype{C}{>{\centering\arraybackslash}m}
\begin{tabular}{|C{22mm}|C{19mm}|C{19mm}|}
\hline
 Orbital inclination & $m_x(M_{\odot})$ & $m_v(M_{\odot})$ \\
\hline
 $i=80^{\circ}$  & 1.84  & 2.6  \\
 $i=88^{\circ}$  & 1.78  & 2.5  \\
\hline
\end{tabular}
\end{table}
\normalsize

Table \ref{old_model} presents the masses of the X-ray pulsar
$m_x$ and optical star $m_v$ corresponding to the minimum $\chi^2$
residual for the theoretical and observed radial-velocity curves.
In Table \ref{old_model}, the component masses are given without
errors, since models of the binary were rejected according to the
selected significance level $\alpha=5\%$. For the quantile
$\Delta(\alpha=5\%)=77.93$, the minimum residual for the orbital
inclinations $i=80^{\circ}$, $i=88^{\circ}$ was
$\chi_{min}^2\simeq170$. Figure \ref{Vr_Roche_old_new} presents
the theoretical radial-velocity curve and Fig.\ref{Xi2}a the
behavior of the $\chi^2$ residual for $i=88^{\circ}$.

\section*{\normalsize\it Results Obtained with Procedure II}

It was noted above, we used the observational data of
\cite{Reynolds1997}. In those spectra, the profile of the $\text
H_{\gamma}$ absorption line at orbital phase 0.47 displays a
distortion [\cite{Reynolds1997}, Fig. 1]. At the same orbital
phase, the theoretical integrated profile of the $\text
H_{\gamma}$ absorption convolved with a response function with
FWHM=1.7\,\AA\, observed profile (Fig. \ref{Hg_FWHM_1_7A_5A}).We
were able to reproduce a similar distortion in the integrated
theoretical profile of the $\text H_{\gamma}$ hydrogen line at
phase 0.47 only with a response-function width of FWHM=5\,\AA\,
(Fig. \ref{Hg_FWHM_1_7A_5A}). Therefore, we carried out our
analysis of the observed radial-velocity curve for the $\text
H_{\gamma}$ hydrogen line with the response-function width
FWHM=5\,\AA.

For the effective temperature of the stellar surface $\sim8000$ K,
the hydrogen line displays a full width at half maximum of the
residual intensity of several tens of \AA. In our case, the width
of the model integrated profile of the $\text H_{\gamma}$ line at
this level was $\sim 25$ \AA\, (Fig. \ref{Hg_Synth_syhth_ph030}).
Therefore, we faced the problem of determining the radial velocity
from such a broad $\text H_{\gamma}$ absorption profile. For
Procedure I, the half-width of the model profile was smaller
$\sim10$ \AA\, (Fig. \ref{Hg_Synth_syhth_ph030}), and the profile
itself was more symmetrical. Therefore, to determine the position
of the center of gravity of the line, it was sufficient to take
three cross sections at heights of 1/3, 1/2, and 2/3 of the
residual intensity. If the integrated $\text H_{\gamma}$ profile
is calculated with Procedure II, three cross sections are not
sufficient, and we selected six and eight levels of the cross
section of the residual intensity of the model integrated profile
to determine the center of gravity of the $\text H_{\gamma}$ line.

\small
\begin{table}[h!]
\caption{Masses of the relativistic and optical components
obtained from the radial velocity determined from six cross
sections of the integrated $\text H_{\gamma}$.}\label{new6_model}
\vspace{3.0mm}\centering
\def\arraystretch{1.4}
\newcolumntype{C}{>{\centering\arraybackslash}m}
\begin{tabular}{|C{22mm}|C{19mm}|C{19mm}|}
\hline
 Orbital inclination & $m_x(M_{\odot})$ & $m_v(M_{\odot})$ \\
\hline
 $i=80^{\circ}$  & $0.85\pm0.15$  & $1.87\pm0.13$  \\
 $i=88^{\circ}$  & $0.81\pm0.13$  & $1.80\pm0.11$  \\
\hline
\end{tabular}
\end{table}
\normalsize

In the former case, the radial velocity was determined from the
wavelength averaged over cross sections at six levels of the lower
half of the absorption profile; the resulting error in the radial
velocities were $\sim1$ km/s. When eight cross-section levels were
used, two cross sections in the upper half of the model profile of
the H¦Ã absorption line were introduced in addition to the six for
the previous case; the resulting error in the theoretical radial
velocities were $\sim2.5$ km/s. Here, we take the radial velocity
at orbital phase $\phi=0.0$ (when the optical component eclipses
the X-ray source) as the error in the theoretical radial velocity.
The error in the radial velocity for the calculation using six
cross sections is appreciably smaller due to the narrowness of the
lower part of the $\text H_{\gamma}$ profile compared to the wings
(Fig. \ref{Hg_Synth_syhth_ph030}).

The mass of the X-ray pulsar $m_x$ and optical star $m_v$
corresponding to the minimum $\chi^2$ residual for the theoretical
and observed radial-velocity curves obtained using six and eight
cross sections of the integrated $\text H_{\gamma}$ line profile
are presented in Tables \ref{new6_model} and \ref{new8_model},
respectively. Figure $\text H_{\gamma}$ presents the
radial-velocity curve obtained using six cross sections of the
integrated \ref{Vr_Roche_old_new} profile.

It is obvious that the results depend on the method used to
determine the center of gravity of the integrated $\text
H_{\gamma}$ absorption profile, although, within the errors, the
mass of the X-ray pulsar is consistent with that of the optical
star.

\small
\begin{table}[h!]
\caption{Masses of the relativistic and optical components
obtained from the radial velocity determined from eight cross
sections of the integrated $\text H_{\gamma}$
profile.}\label{new8_model} \vspace{3.0mm}\centering
\def\arraystretch{1.4}
\newcolumntype{C}{>{\centering\arraybackslash}m}
\begin{tabular}{|C{22mm}|C{19mm}|C{19mm}|}
\hline
 Orbital inclination & $m_x(M_{\odot})$ & $m_v(M_{\odot})$ \\
\hline
 $i=80^{\circ}$  & 0.70  & 1.75  \\
 $i=88^{\circ}$  & 0.70  & 1.70  \\
\hline
\end{tabular}
\end{table}
\normalsize

The model of the binary obtained using six cross sections is
consistent with the observational data at the $\gamma=95$\%
confidence level, with the minimum residual $\chi_{min}^2\simeq70$
(Fig.\ref{Xi2}b). When eight cross sections are used, the binary
model is rejected at the significance level $\alpha=5\%$, with the
minimum residual $\chi_{min}^2\simeq90$ (for this reason, the
component masses in Table \ref{new8_model} are presented without
errors).

\section*{\normalsize\it Integrated Profile and the Radial-velocity Curve
Including non-LTE Effects}

As is noted above, in spite of the substantial X-ray heating
$(k_x=150)$, the absorption lines in the optical spectrum of the
HZ Her system display no visible emission components. This is
apparently due to non- LTE effects (see above).

\begin{figure*}
\vspace{0cm} \epsfxsize=0.99\textwidth
\epsfbox{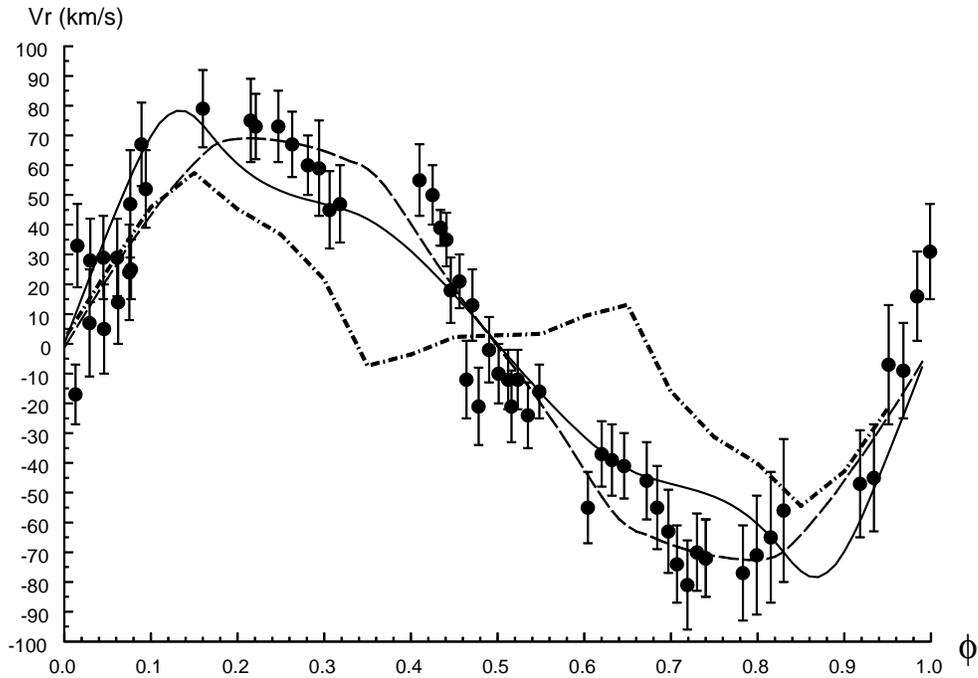}
\caption{Observed radial-velocity curve for the optical component
of the X-ray binary Her X-1 taken from \cite{Reynolds1997}
(points) and theoretical radial-velocity curves calculated with
Procedure I for $m_x=1.78M_{\odot}$, $m_v=2.5M_{\odot}$, and
$i=88^{\circ}$ (solid curve); and with Procedure II for
$m_x=0.81M_{\odot}$, $m_v=1.80M_{\odot}$, and $i=88^{\circ}$ from
six cross sections of the integrated profile of the $\text
H_{\gamma}$ line (dashed curve) and using the $NONLTE3$
\cite{sahi2} package in the non-LTE model for $m_x=0.81M_{\odot}$,
$m_v=1.80M_{\odot}$, $i=88^{\circ}$. (dash-dotted curve).}
\label{Vr_Roche_old_new}
\end{figure*}

\begin{figure*}
\vspace{0cm} \epsfxsize=0.99\textwidth
\epsfbox{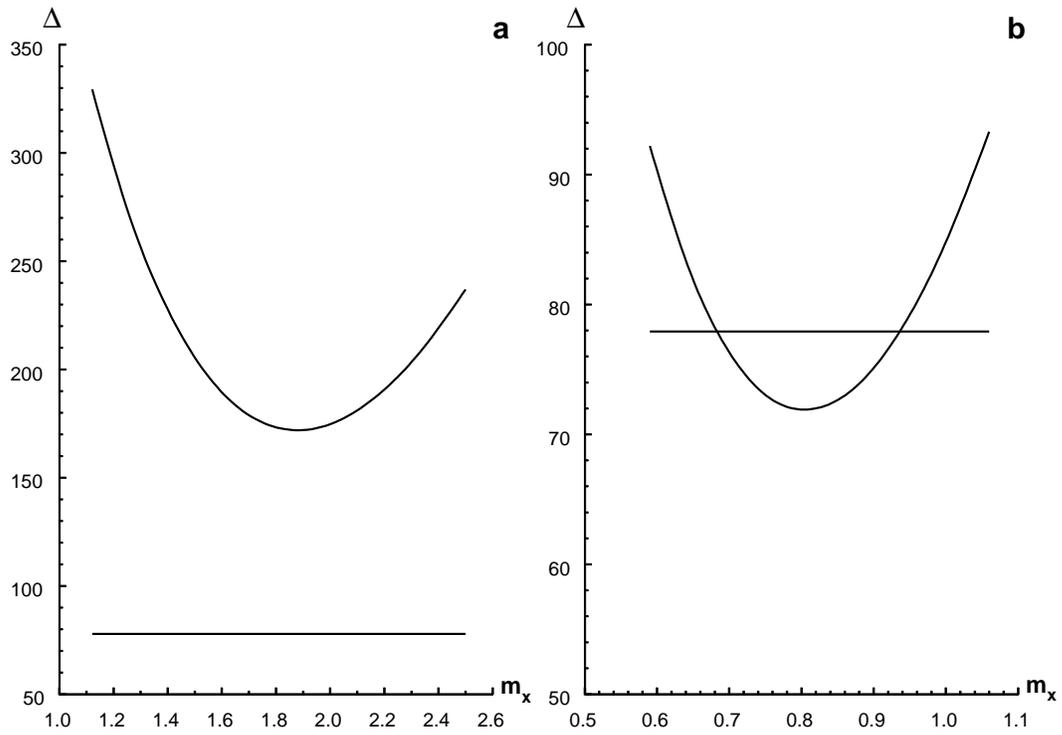}
\caption{Discrepancies between the observed and theoretical radial
velocity curves of HZ Her. The theoretical curves were obtained
with (a) Procedure I and (b) Procedure II, for the orbital
inclination $i=80^{\circ}$. The horizontal line corresponds to the
critical level of the $\chi^2$ criterion $\Delta_{59}=77.93$ for
the significance level $\alpha=5$\%.} \label{Xi2}
\end{figure*}

\begin{figure*}
\vspace{0cm} \epsfxsize=0.99\textwidth
\epsfbox{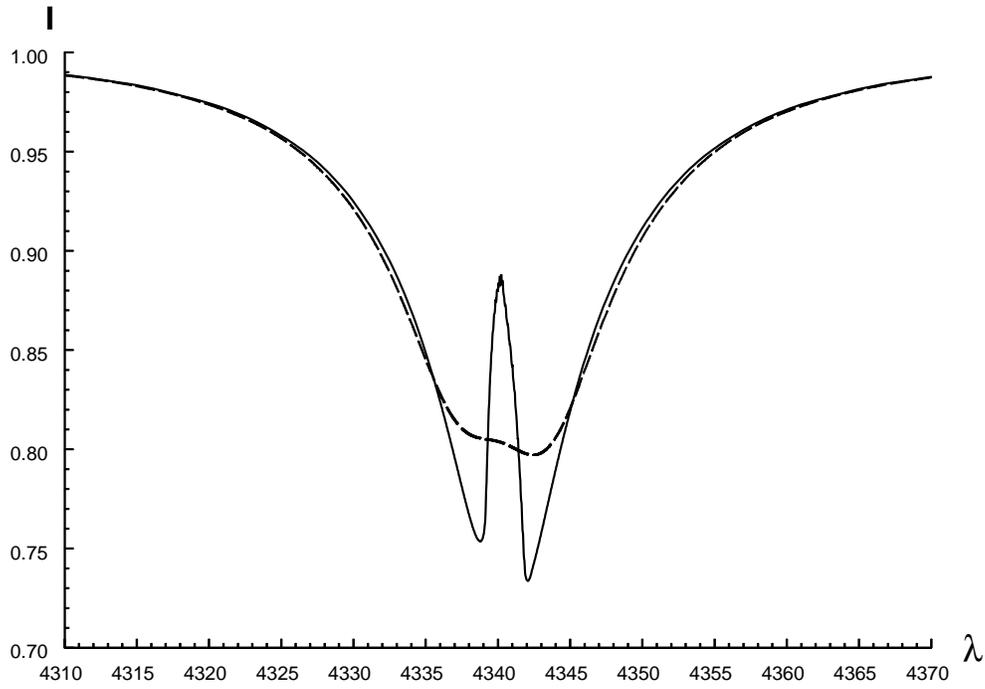}
\caption{Model integrated profile of the hydrogen absorption line
at orbital phase $\phi=0.47$, calculated assuming LTE with
Procedure II for the response-function widths FWHM=1.7\,\AA,(solid
curve) and FWHM=5\,\AA, (dashed curve) for $m_x=0.81M_{\odot}$,
$m_v=1.80M_{\odot}$, and $i=88^{\circ}$.}\label{Hg_FWHM_1_7A_5A}
\end{figure*}

\begin{figure*}
\vspace{0cm} \epsfxsize=0.99\textwidth
\epsfbox{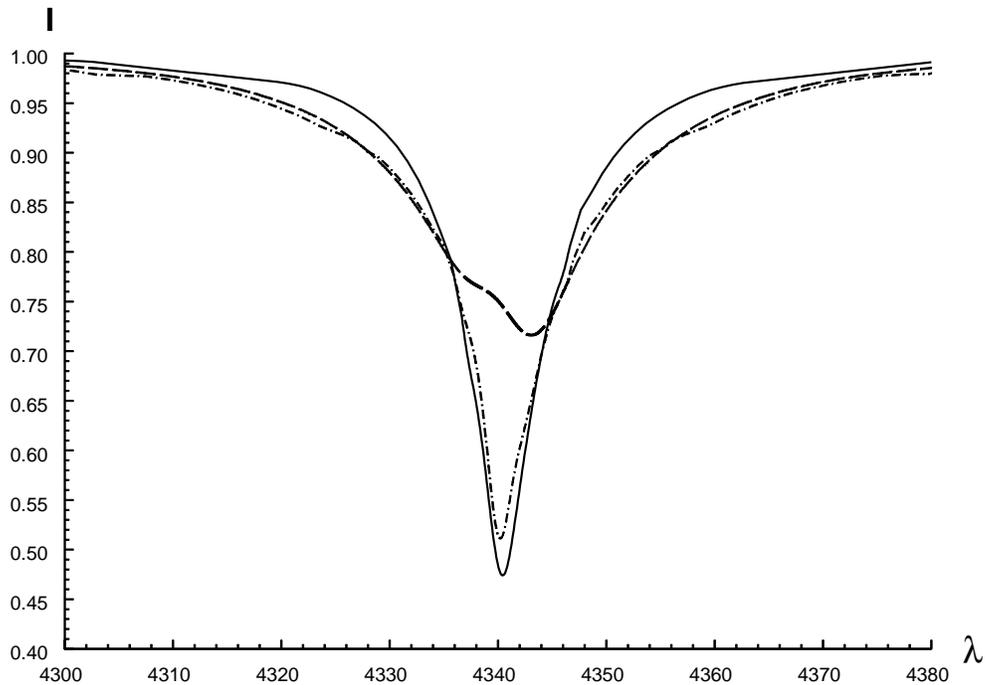}
\caption{Model integrated profile of the $\text H_{\gamma}$
hydrogen absorption line at orbital phase $\phi=0.30$, obtained
with Procedure II for the response-function width FWHM=5\,\AA, and
$m_x=0.81M_{\odot}$, $m_v=1.80M_{\odot}$, and $i=88^{\circ}$
(dashed curve); with Procedure I for the response-function width
FWHM=1.7\,\AA, and $m_x=1.78M_{\odot}$, $m_v=2.5M_{\odot}$, and
$i=88^{\circ}$ (solid curve); and with the $NONLTE3$ package in
the non-LTE model for the response-function width FWHM=1.7\,\AA,
and $m_x=0.81M_{\odot}$, $m_v=1.80M_{\odot}$, and $i=88^{\circ}$
(dash-dotted curve).} \label{Hg_Synth_syhth_ph030}
\end{figure*}

We also calculated the integrated $\text H_{\gamma}$ absorption
profiles and corresponding radial-velocity curve in a non-LTE
model (Fig. \ref{Vr_Roche_old_new}). These calculations were
performed using the NONLTE3 code \cite{sahi2} for the model binary
parameters $m_x=0.81M_{\odot}$, $m_v=1.80M_{\odot}$, and
$i=88^{\circ}$. (see the remaining parameters in Table \ref{param
Roche}), taking the width of the response function to be
FWHM=1.7\,\AA. Regrettably, calculations with the $NONLTE3$
procedure are time consuming, which currently makes searches for
the parameters of close binaries challenging, and we restrict our
consideration here to qualitative conclusions.

Figures \ref{Hg_loc} and \ref{Hg_Synth_syhth_ph030} present the
local and integrated profiles of the $\text H_{\gamma}$ absorption
line calculated in the non-LTE approximation. Figure
\ref{Vr_Roche_old_new} presents the radial-velocity curve derived
from these profiles.

We can see from Fig.\ref{Hg_Synth_syhth_ph030} that,
qualitatively, the integrated $\text H_{\gamma}$ absorption
profile obtained in the non-LTE model is close to the integrated
profile calculated using Procedure I - in both profiles, no
emission component is seen. At the same time, the integrated
$\text H_{\gamma}$ absorption profile obtained assuming LTE with
Procedure II displays an emission component (see
Figs.\ref{Hg_SpXray} and \ref{Hg_Synth_syhth_ph030}).

We can also see in Fig.\ref{Vr_Roche_old_new}  that the shape of
the radial-velocity curve obtained with the non- LTE model is
qualitatively closest to that calculated with Procedure I. In this
connection, the results of fitting the observed radial-velocity
curve for HZ Her/Her X-1 made with Procedure I may bemore
reliable: $m_x=1.78M_{\odot}$, $m_v=2.5M_{\odot}$ for the orbital
inclination $i=88^{\circ}$. We stress again, however, that the
model of the binary system with these parameters was rejected at
the $\alpha=5\%$ significance level adopted in the study.

\section*{\normalsize DISCUSSION}

Let us consider the advantages and disadvantages of the procedures
used to calculate the radial-velocity curves in our study.

When the integrated profile of the optical component is calculated
with Procedure I, we use integrated profiles of absorption lines
of single stars tabulated by Kurucz \cite{Kurucz1979} for the
local profiles. These absorption profiles do not contain emission
components, and the local profiles used in Procedure I are
qualitatively consistent with the local profiles of the $\text
H_{\gamma}$ absorption lines in the non-LTE approximation.
However, the model integrated $\text H_{\gamma}$ profile obtained
with Procedure I does not reproduce the distinctive distortion of
the $\text H_{\gamma}$ absorption line at orbital phase
$\phi=0.47$ (\cite{Reynolds1997}, Fig. 1), and the best-fit
theoretical radial-velocity curve seems to diverge from the
observed radial-velocity curve both qualitatively and
quantitatively ($\chi_{min}^2\simeq170$ - the model is rejected at
the $\alpha=5$\% significance level; Fig.\ref{Vr_Roche_old_new}).
In this light, we cannot take our estimates for the Her X-1/HZ Her
component masses obtained with Procedure I to be final.

Let us consider the results of our fitting of the observational
data and the specific features of calculating the model absorption
profiles with Procedure II.

The local profiles assuming LTE with exact account of the effect
of heating contain an emission component (which is not seen in the
observed optical spectra of HZ Her), whereas the local $\text
H_{\gamma}$ absorption profiles calculated in the non-LTE
approximation display no emission component (Fig.\ref{Hg_loc}).
Nonetheless, the theoretical integrated $\text H_{\gamma}$
absorption profile convolved with an artificially broadened
response function with FWHM=5\,\AA\, (exceeding the real width of
the response function by a factor of three) qualitatively
reproduce the "distortion"\, of the observed absorption profile
shape (cf. Fig. 1 in \cite{Reynolds1997} and
Fig.\ref{Hg_Synth_syhth_ph030}). The full width of the obtained
integrated $\text H_{\gamma}$ absorption profile at half the
residual intensity was $\sim 25\text \AA$, hindering determination
of the center of gravity of the absorption line. Six or eight
cross sections were used to determine the center of gravity of the
model integrated $\text H_{\gamma}$ absorption profile.

The component masses obtained with six and eight cross sections of
the model integrated $\text H_{\gamma}$ profile are consistent
within the errors (Tables \ref{new6_model} and \ref{new8_model}).
However, both the quantitative estimates of the mass of the binary
components obtained with Procedure II and the reliability of the
binary model depend on the number of cross sections of the
integrated profiles used to determine the center of gravity of the
$\text H_{\gamma}$ absorption profile. For example, when eight
cross sections of the model integrated $\text H_{\gamma}$ profile
were used, the model was rejected at selected significance level
of $\alpha=5\%$, and the masses for the binary components differed
by $\sim 0.1M_{\odot}$ from those obtained when six cross sections
of the integrated profile were used.

Thus, it does not seem reasonable to give preference to any one of
the various results, and we consider the results presented in
Tables \ref{old_model} and \ref{new6_model} equally valid. As we
have shown here, reliable estimation of the component masses for
the Her X-1/HZ Her binary system is possible only based on an
analysis of the radial-velocity curve jointly with the orbital
variability of the absorption profiles, with correct account for
non-LTE effects.

\section*{\normalsize CONCLUSION}

Our study has mainly focused on the methodical side of the problem
considered. We have shown that determining the radial velocities
in close binaries with intense X-ray heating of the optical star
requires taking into account non-LTE effects in the formation of
absorption lines in the spectrum of the optical star. For Her
X-1/HZ Her-type binary systems, not only the radial-velocity curve
(which is specified by the line profiles only ambiguously)must be
analyzed, but also line profiles at different phases of the
orbital period.

Our calculations have shown that, due to the very intense X-ray
heating in Her X-1/HZ Her, the $\text H_{\gamma}$ absorption line
of the optical star is appreciably asymmetrical, making it
difficult to determine unambiguously the center of gravity of its
profiles and, as a consequence, the radial velocities and
component masses. For example, when the theoretical radial
velocity is determined using different numbers of cross sections
of the absorption-line profile, the mass of the X-ray pulsar $m_x$
varies by $\sim15$\% (Tables \ref{new6_model} and
\ref{new8_model}).We stress that the methods for estimating the
Her X-1/HZ Her component masses presented here are based on the
most current and physically justified techniques, in which local
absorption-line profiles are calculated taking into account
external X-ray heating. Nonetheless, even this approach cannot
yield unambiguous mass estimates for the X-ray pulsar in Her
X-1/HZ Her, due to the enormous rate of X-ray heating in this
unique X-ray binary. Analyzing the radial-velocity curve in two
ways, we have obtained two masses for the X-ray pulsar:
$m_x=0.85\pm0.15M_{\odot}$ and $m_x=1.8M_{\odot}$. Thus, the mass
of the X-ray pulsar in Her X-1 is currently determined only to
within a factor of $\sim2$, due to uncertainty in realistic models
for the formation of line profiles in the spectrum of the heated
optical star.

Our results demonstrate the need to use high-resolution
spectrograms ($\lambda/\Delta\lambda\simeq50000$) and to take into
account non-LTE effects when estimating the mass of the compact
object. Correct and reliable estimates for the component masses
can be obtained only by analyzing also the orbital variability of
the absorption-line profiles in the spectrum of the optical
component in a non-LTE approximation.

We thank I.I. Antokhin for useful discussions.

\renewcommand{\bibname}{References}

\newpage

\end{document}